\documentclass[sigconf]{acmart}
\usepackage{balance}
\usepackage{amsmath}
\usepackage{threeparttable}
\usepackage{natbib} 
\newcommand{\mycomment}[1]{}
\AtBeginDocument{%
  }

\copyrightyear{2025}
\acmYear{2025}
\setcopyright{acmlicensed}
\acmConference[WWW Companion '25]{Companion
Proceedings of the ACM Web Conference 2025}{April 28-May 2, 2025}{Sydney,
NSW, Australia}
\acmBooktitle{Companion Proceedings of the ACM Web Conference 2025 (WWW
Companion '25), April 28-May 2, 2025, Sydney, NSW, Australia}
\acmDOI{10.1145/3701716.3715242}
\acmISBN{979-8-4007-1331-6/2025/04}

\begin{document}

\title[Knowledge Distillation for Enhancing Walmart E-commerce Search Relevance]{Knowledge Distillation for Enhancing Walmart E-commerce Search Relevance Using Large Language Models}


\author{Hongwei Shang}
\authornote{Both authors contributed equally to this research.}
\email{hongwei.shang@walmart.com}
\authornote{Corresponding author.}
\orcid{0009-0005-9856-9178}
\affiliation{
  \institution{Walmart Global Tech}
   \city{Sunnyvale}
   \state{CA}
   \country{USA}
}

\author{Nguyen Vo}
\authornotemark[1]
\email{vknguyen09@gmail.com}
\orcid{0000-0001-5888-4740}
\affiliation{
  \institution{Amazon Inc.}
   \city{Sunnyvale}
   \state{CA}
   \country{USA}
}

\author{Nitin Yadav}
\email{nitin.yadav@walmart.com}
\orcid{0009-0003-8051-9741}
\affiliation{
  \institution{Walmart Global Tech}
   \city{Sunnyvale}
   \state{CA}
   \country{USA}
}

\author{Tian Zhang}
\email{tian.zhang@walmart.com}
\orcid{0009-0009-8198-9060}
\affiliation{
  \institution{Walmart Global Tech}
   \city{Sunnyvale}
   \state{CA}
   \country{USA}
}

\author{Ajit Puthenputhussery}
\email{ajit.puthenputhussery@walmart.com}
\orcid{0000-0001-7141-1534}
\affiliation{
  \institution{Walmart Global Tech}
   \city{Sunnyvale}
   \state{CA}
   \country{USA}
}

\author{Xunfan Cai}
\email{xunfan.cai@walmart.com}
\orcid{0009-0001-3973-8577}
\affiliation{
  \institution{Walmart Global Tech}
   \city{Sunnyvale}
   \state{CA}
   \country{USA}
}

\author{Shuyi Chen}
\email{shuyi.chen@walmart.com}
\orcid{0009-0005-5982-0235}
\affiliation{
  \institution{Walmart Global Tech}
   \city{Sunnyvale}
   \state{CA}
   \country{USA}
}

\author{Prijith Chandran}
\email{prijith.chandran@walmart.com}
\orcid{0009-0001-4598-5064}
\affiliation{
  \institution{Walmart Global Tech}
   \city{Sunnyvale}
   \state{CA}
   \country{USA}
}

\author{Changsung Kang}
\email{changsung.kang@walmart.com}
\orcid{0009-0007-5305-8256}
\affiliation{
  \institution{Walmart Global Tech}
   \city{Sunnyvale}
   \state{CA}
   \country{USA}
}

\renewcommand{\shortauthors}{Hongwei Shang et al.}

\begin{abstract}
Ensuring the products displayed in e-commerce search results are relevant to users' queries is crucial for improving the user experience. 
With their advanced semantic understanding, deep learning models have been widely used for relevance matching in search tasks.
While large language models (LLMs) offer superior ranking capabilities, it is challenging to deploy LLMs in real-time systems due to the high-latency requirements. 
To leverage the ranking power of LLMs while meeting the low-latency demands of production systems,
we propose a novel framework that distills a high-performing LLM into a more efficient, low-latency student model.
To help the student model learn more effectively from the teacher model,
we first train the teacher LLM as a classification model with soft targets.
Then, we train the student model to capture the relevance margin between pairs of products for a given query using mean squared error loss.
Instead of using the same training data as the teacher model,
we significantly expand the student model’s dataset by generating unlabeled data and labeling it with the teacher model’s predictions. 
Experimental results show that the student model's performance continues to improve as the size of the augmented training data increases. 
In fact, with enough augmented data, the student model can outperform the teacher model. 
The student model has been successfully deployed in production at Walmart.com with significantly positive metrics.

\end{abstract}



\maketitle

\section{Introduction}

In recent years, there has been exceptionally rapid growth in e-commerce platforms.
On e-commerce platforms such as Walmart, Amazon and Taobao, customers enter search queries, 
and the platform must deliver the most relevant items from millions of products in the product catalog.
As the product catalog continues to grow, continuously improving search ranking systems has become essential for showing customers highly relevant items on e-commerce platforms.
Product search ranking engine typically comprises two main stages: \textbf{the retrieval stage}
where a set of relevant products candidates is retrieved from the product catalog to form the recall set,
\textbf{the rerank stage} where the candidates obtained from the previous stage are re-ranked to 
form a ranked product list to be shown to customers.

The embedding-based two tower model has become an important approach in the retrieval stage of
e-commerce search engines, especially the bi-encoder which uses transformer-encoded representations.
Bi-encoder models learn embedding representations for search queries and products from training data.
At retrieval time, semantically similar items are retrieved based on simple similarity 
metrics, such as the cosine similarity between the query and the item. 
Since the query and document embeddings are learned through separate models, 
this approach allows the embeddings to be precomputed offline with minimal computational overhead and low latency \citep{yates2021pretrained,bai2022improving}.
In contrast, the cross-encoder model takes the query and item text as input and directly outputs a prediction.

Although bi-encoder approaches are typically much faster, they are generally less effective than cross-encoder methods \citep{Vo:Shan:2024:Know,Wang:Sund:2024:Impr}.
By concatenating query and item as input, the cross-encoder model can benefit from the attention mechanism across all tokens in the inputs,
thus capturing interactions between query and item text.
As a result, the trade-off with bi-encoders involves sacrificing the model's effectiveness for large latency gain. 
The retrieval stage has strict latency requirements as it needs to search through millions of products in the product catalog.
However, the downstream re-rank stage only ranks the products retrieved during the retrieval stage,
allowing it to afford more relaxed latency requirements compared to the retrieval stage. 

Our previous production model is the first ranking model with a cross-encoder (XE) feature that can better
understand the relevance of items to search queries.
In this work, we present our our enhanced BERT-Base ($BERT_{\text{Base}}$) model with
cross-encoder architecture,
optimized through knowledge distillation from LLMs on an augmented large-scale training dataset.

In this work, our goal is to improve effectiveness of cross-encoder models used in production for the tail segment.
To achieve this goal, we propose a novel knowledge distillation framework that distills the teacher LLM model
(i.e. 7 billion parameters) into a student model (i.e. $BERT_{\text{Base}}$ \citep{kenton2019bert}) 
with significantly augmented training data. 
This framework improves the student model's effectiveness while preserving its efficiency. 
First, we train the LLM teacher model using soft-label targets converted from human editorial labels.
In addition to item's title, product type, brand, color and gender, 
we further incorporate item's description into the model input
to improve our teacher LLM model's performance. 
After that, the student model is trained to mimic the margin between two documents predicted by the teacher model. 
Rather than distilling the knowledge on the labeled training data, 
we generate large-scale unlabeled data from user's log data,
and label it with the teacher model's predictions such that the student model can learn from the teacher at a much larger dataset. 
The effectiveness of our cross-encoder based student model is demonstrated through both offline and online 
experiments, showing its ability to improve both the relevance and engagement metrics of the system.
Our cross-encoder based student model has been deployed
in Oct 2024 as the most dominant feature in the current production system, and is powering tail queries in Walmart.com's e-commerce engine.

Our main contributions are summarized as follows:
\begin{itemize}

\item We propose a knowledge distillation framework in which the teacher model is trained with soft-labeling, enabling more effective distillation through the teacher's single output. 
The distillation process uses a extended Margin-MSE loss on an augmented, large-scale unlabeled dataset.
Building on the work of \citet{hofstatter2020improving}, 
where both the student and teacher models share the same training data and require true labels for distillation, 
we significantly expand the dataset used to train the student model. 
Specifically, we generate a large-scale unlabeled dataset 
and use the teacher model's predictions to label the data. 
The Margin-MSE loss is then computed on all item pairs for a given query, 
eliminating the need for true labels.

\item We conduct an ablation study that reveals the following findings: 
(1) using Margin-MSE loss to align the margin with the teacher model significantly improves knowledge distillation compared to using pointwise cross-entropy (CE) loss.
(2) distilling knowledge through a large-scale dataset labeled with the teacher's predictions leads to a substantial performance boost, enabling the student model to achieve results comparable to the teacher model.
(3) the student model performs on par with the teacher model, even without the `item description' field, which has been shown to be beneficial. 
These results demonstrate that increasing the model's capacity is not necessary to effectively learn from the teacher model.

\item Our proposed model has been successfully deployed in production at Walmart.com with significantly positive metrics.
\end{itemize}

The rest of this paper is organized as follows: 
in Section~\ref{sec:relatework}, we review related work and discuss the previous production model.
Section~\ref{sec:architecture} outlines our model architecture and loss functions.
In Section~\ref{sec:experiments}, we present our offline experiment setup and evaluation metrics, followed by the online test results in Section~\ref{sec:online_experiment}.
Finally, Section~\ref{sec:conclusion} concludes with a discussion of future work.

\section{Related Work}
\label{sec:relatework}
Search ranking has long been a popular topic within the search community.
Traditional method in search ranking are based on text match between queries
and items like LSI \citep{deerwester1990indexing}, BM25 \citep{robertson2009probabilistic}
and its variations \citep{Trot:Puur:Impr:BM25:2014:ADCS,Tan:Jiang:StatBM25:2018:ICTIR}.
They struggle with handling the lexical gap between queries and product descriptions. 
Neural network models can learn representations of queries and
documents from raw text, helping to bridge the lexical gap between the query and documents vocabularies
to some extent.
Many large e-commerce platforms have deployed neural network models in their product search engines
e.g. Walmart \citep{magnani2022semantic,peng2023entity}, Amazon \citep{nigam2019semantic}, 
Facebook Marketplace \citep{he2023que2engage}, 
Taobao \citep{zheng2022multi}, JD.com \citep{wang2023learning}, 
Baidu \citep{hager2024unbiased}. 
Recent advancements in e-commerce search have focused on various aspects, 
including retrieval, ranking, and query understanding systems.
\citet{lin2024enhancing} addressed relevance degradation in Walmart's retrieval system with a relevance reward model, while \citet{peng2023entity} introduced an entity-aware model to enrich query representation with engagement data. \citet{luo2024exploring} proposed a multitask learning framework for ranking optimization at Amazon, and \citet{peng2024large} developed a query rewriting framework to address the semantic gap in long-tail queries.

Recent advances in LLMs have revolutionized semantic search,
with fine-tuned LLMs achieving accuracy close to human performance on labeled datasets \citep{Toma:Spie:2024:Larg,Mehr:Moha:Larg:2024}.
However, due to latency constraints, LLMs cannot be deployed 
for real-time inference in production.
Knowledge distillation \citep{hinton2015distilling} bridges this gap 
by transferring LLM capabilities to lighter, deployable models.
Knowledge distillation methods can be mainly categorized into three types:
response-based \citep{hofstatter2020improving,menon2022defense,he2022metric,liu2022knowledge,ye2024multilingual,ye2022multilingual}, 
representation-based \citep{yao2022reprbert,liu2023multimodal}, 
and relation-based methods \citep{liu2019knowledge}. 
Specifically, \citet{hofstatter2020improving} introduced a margin focused loss (Margin-MSE)
to adapt knowledge distillation across different architectures.
\citet{sun2023instruction} introduced a novel instruction distillation method 
to improve the efficiency of LLMs by converting complex pairwise ranking 
into more efficient pointwise ranking. 
Building on the work of \citet{hofstatter2020improving},
we expand the student model’s dataset by incorporating augmented unlabeled data, 
which we label using the teacher model’s predictions. 
Our knowledge distillation approach does not require manual labeling of query-item pairs
for the student model.

\paragraph{Previous Production Model}

\begin{figure}
  \centering
     \includegraphics[width=1.0\linewidth]{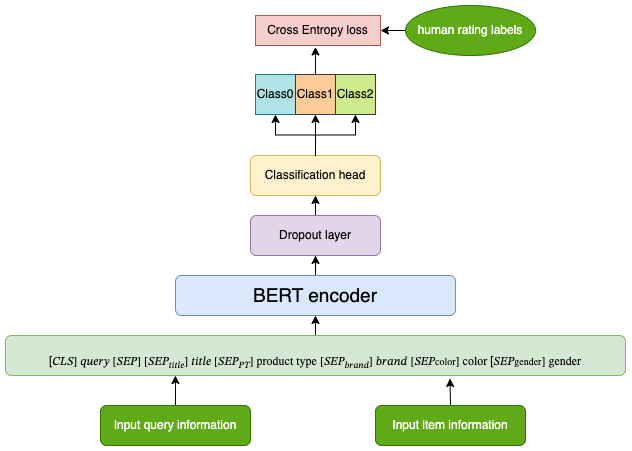} \\
  \caption{The architecture for the previous production version of the Cross encoder}
  \label{fig:xe_existing_prod}
\end{figure}

The previous BERT-based cross-encoder model treats the classification task as a multi-class classification problem, outputting probabilities over three classes (class 0, class 1, and class 2). 
It uses a Walmart in-house pre-trained BERT model as the starting checkpoint, 
which is then fine-tuned on human-labeled query-item pairs (QIPs) (human-judged\_v1 in 

~\ref{tab:datastat}).
It aims to classify the relevance of these pairs. 
As shown in Figure~\ref{fig:xe_existing_prod}, 
the query and item text are concatenated into a single string and fed into the BERT encoder. After encoding, non-linear layers transform the embedding into a three-dimensional classification vector, which is then passed through a softmax layer to generate a three-class probability vector. This vector is optimized using cross-entropy loss.

The outputs of the cross-encoder model are used as features in our tail relevance model, which ranks results for the tail queries. 
While the launch of the cross-encoder in April 2024 marked a significant improvement over the previous model (which lacked the cross-encoder feature), 
there is still substantial room for further enhancement in relevance for long-tail queries.
One limitation of the previous model was the insufficient training data, primarily due to the high costs associated with human annotators' labeling. Additionally, the model suffered from a lack of sufficient negative-label examples and the presence of incorrect labels in the training data.

\section{Architecture}
\label{sec:architecture}
\subsection{Problem Formulation}

\begin{table}
\caption{Example of a query and items with class 2, class 1 and class 0 respectively}
\centering
\label{tab:rel_example}
\begin{tabular}{rrr}
    \toprule
Query & Item & Class label \\
\midrule 
apple iphone 14 pro max & iphone 14 pro max & 2 \\
apple iphone 14 pro max & iphone 14 pro & 1 \\
apple iphone 14 pro max & iphone 14 pro max case & 0 \\
    \bottomrule
\end{tabular}
\end{table}

Given a query $q$ and an item $d$, where each item $d$ has attributes 
such as title, product type, brand, color, gender and description, 
our goal is to train a teacher model using LLM,
and then distill this knowledge into a more lightweight student model 
using a large-scale, augmented dataset. 
We denote the predicted relevance for query $q$ and document $d$
as $t(q,d) \in \mathcal{R}$ from the teacher model,
and $s(q,d) \in \mathcal{R}$ from the student model.

\begin{table}
\caption{Rating score and relevance label mapping}
\centering
\label{tab:rel_label}
\begin{tabular}{lllll} 
    \toprule
Rating  & Description & Class & Soft \\
Score & & Label & Target \\
\hline
4 & Excellent - perfect match & 2 & 1\\
3 & Good - item with a mismatched attribute & 1 & 0.5 \\
2 & Okay & 0 & 0\\
1 & bad - Irrelevant & 0 & 0 \\
0 & Embarrassing & 0 & 0 \\
    \bottomrule
\end{tabular}
\end{table}

\begin{figure*}
  \centering
    \includegraphics[width=1.0\linewidth]{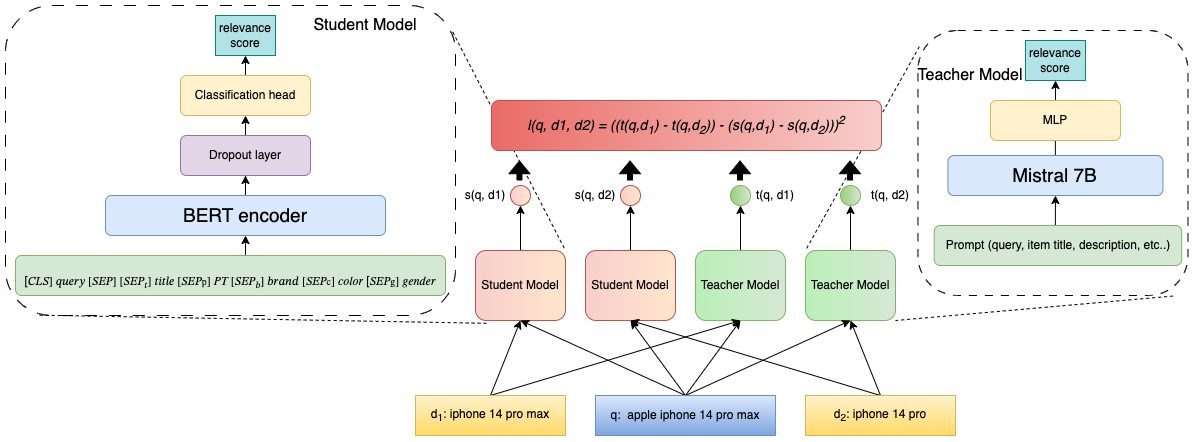}
  \caption{Our proposed knowledge distillation framework}
  \label{fig:kd_figure}
\end{figure*}

Our guidelines categorize each $(q,d)$ pair into five ordered ratings, 4, 3, 2, 1, and 0. 
These ratings are then mapped to three classes (Class 2, Class 1, Class 0) based on our production requirements, as shown in Table~\ref{tab:rel_label}. 
Examples of these three relevance classes are provided in Table~\ref{tab:rel_example}.

In this work, we formulate search relevance prediction as a binary classification problem with soft targets \citep{Vo:Shan:2024:Know}. 
We map the three-class labels into soft targets (as shown in Table~\ref{tab:rel_label}),
where an excellent item is labeled as 1, a good item as 0.5, and an irrelevant item as 0.
The advantage of predicting a single output is that it enhances the effectiveness of knowledge distillation. 
Our framework consists of two main components: (1) the cross-encoder architecture based LLM used as the teacher model, (2) a more lightweighted cross-encoder architecture model serving as the student model. 
First, we fine-tuned the LLM as a teacher model using human editorial labels. 
Then, we use the teacher model to make inferences on a large-scale, 
augmented set of unlabeled query-item pairs obtained from search log data,
generating a substantial training data for the student model. 
The details of these components are provided in the following subsections. 

\subsection{LLM as the Teacher Model}
\label{sec:LLM}

We use LLMs to predict the relevance of an item to a query based on textual information. 
For training our teacher model, we employ two 7B-parameter open-source LLMs: 
Llama2-7B \citep{touvron2023llama} and Mistral-7B-v0.1 \citep{jiang2023mistral}.
The model training is implemented using PyTorch \citep{paszke2019pytorch} 
and the Huggingface Transformers framework \citep{wolf2019huggingface}. 
Both Llama2-7B and Mistral-7B-v0.1 are fine-tuned using a cross-encoder architecture 
to predict the relevance between a query and an item. 
To improve training efficiency, we apply the softRank Adaptation (LoRA) \citep{Hu:Shen:2022:Lora}, 
which has been shown to outperform full model fine-tuning when trained on datasets of size
on the order of $10^6$ \citep{Mehr:Moha:Larg:2024}.

Our teacher models are trained as classification models, 
using concatenated query and item textual features as inputs, 
along with soft relevance labels.
The classification is based on the ``last token'', 
where the hidden state of this token is used as the representation of the query-item pair.
An MLP layer is then applied to transform this representation into a single output value, 
which represents the predicted relevance score $t(q,d)$ of the teacher model. 
We use cross-entropy loss to train the LLM, as defined by $\mathcal{L}_{teacher}(q)$
for each query $q$: 
\begin{equation*}
   \mathcal{L}_{teacher}(q) = \sum_{d \in \mathbf{D}_q}[-y_{q,d} \cdot \log t(q,d)-(1-y_{q,d}) \cdot \log(1 - t(q,d))],
\end{equation*}
where $\mathbf{D}_q$ is the document collection for query $q$,
and $y_{q,d} \in \{0, 0.5, 1\}$ is the soft label for query $q$ and document $d$,
converted from the original editorial feedback. 
The objective is to optimize $t(q,d)$ by minimizing the above loss function. 

Note that the teacher model incorporates additional text data, specifically the ``item description'', 
which has been shown to improve performance in multi-billion parameter LLMs \citep{Mehr:Moha:Larg:2024}. According to \citet{Mehr:Moha:Larg:2024}, one of the key advantages of LLMs is their ability to effectively process long item descriptions—information that is often difficult for human evaluators to review quickly.

\subsection{Cross-encoder Architecture based Student Model}
\label{sec:student}

Our student model is built on a $BERT_{\text{Base}}$ cross-encoder architecture.
Similar to the teacher model, we concatenate the query with the item's attributes,
including title, product type (PT), brand, gender and color. 
To meet production latency requirements, we exclude the item description from the input.
Each item attribute is separated by unique separator tokens, as shown below:
$$
   {\scriptstyle [CLS]\; query \;[SEP] \; [SEP_t] title [SEP_p] PT [SEP_{b}] brand [SEP_c] color [SEP_g] gender}
$$,
where $[SEP_t]$, $[SEP_p]$ and $[SEP_b]$, $[SEP_c]$ and $[SEP_g]$ are 
distinct $[unused]$ tokens from the BERT vocabulary. 
After the encoder pass, the hidden state of $[CLS]$ token is used
as the representation of the query-item pair. 
An MLP layer is then applied to this hidden state to generate a single prediction score. 

Our student model training is based on a loss function similar to the margin MSE loss \citep{hofstatter2020improving,Vo:Shan:2024:Know}. 
Based on the observation that different architectures converge to different scoring ranges, 
\citep{hofstatter2020improving} proposed distilling knowledge 
not by directly optimizing the raw scores for each query-item pair, 
but by focusing on the margin between relevant and non-relevant documents
for each query, using a Margin-MSE loss. 
The relevance of a document is determined by its true label. 
In their work, the training data for the student model is the same as that for the teacher model, 
meaning that all query-item pairs have known true labels (either relevant or non-relevant) 
for knowledge distillation.
We extend this approach by augmenting the training data with a much larger set of unlabeled data,
where the true labels are unknown. 
Building on the works of \citet{hofstatter2020improving} and \citet{Vo:Shan:2024:Know} which 
focus on distilling knowledge from the teacher model using labeled data, 
we adapt the approach to handle unlabeled data. 
Specifically, we compute the Margin-MSE loss for all triplets $(q, d_i, d_j)$,
where $d_i$ and $d_j$ are documents from the document collection for query $q$, 
denoted by $\mathbf{D}_q$.
Let  $|\mathbf{D}_q|$ denote the number of documents in $\mathbf{D}_q$, 
which results in $|\mathbf{D}_q|(|\mathbf{D}_q|-1)/2$ triplets, 
as shown in Equation~(\ref{eq:margin_mse}).

In contrast to \citet{hofstatter2020improving}, who sample triplets $(q, d^+, d^-)$ 
where $d^+$ is a relevant document and $d^-$ is an irrelevant document 
based on the training data's relevance labels, 
our approach relies entirely on the teacher model's predictions 
and does not require any labeling of the query-item pairs. 
This extension makes knowledge distillation feasible for large-scale augmented datasets, 
even when they are unlabeled.
Let $\Delta^t_{q,d_i,d_j}$ and $\Delta^s_{q,d_i,d_j}$ represent
the score differences for the pairs $(q,d_i)$ and $(q,d_j)$
from the teacher and student models respectively, where 
$\Delta^t_{q,d_i,d_j} = t(q,d_i) - t(q, d_j)$
and 
$\Delta^s_{q,d_i,d_j} = s(q,d_i) - s(q, d_j)$. 
The goal of the student model is to make $\Delta^s_{q,d_i,d_j}$ as close as possible to $\Delta^t_{q,d_i,d_j}$.
The loss function for the student model for query $q$ is defined as:
\begin{equation}
\mathcal{L}_{student}(q) = 
\frac{2}{ |\mathbf{D}_q| ( |\mathbf{D}_q|-1)}
\sum_{\substack{d_i, d_j \in \mathbf{D}_q, \\ 1 \leq i < j \leq |\mathbf{D}_q|}} (\Delta^t_{q,d_i,d_j} -  \Delta^s_{q,d_i,d_j})^2. 
\label{eq:margin_mse}
\end{equation}

\begin{table*}[ht]
    \caption{Statistics of Training datasets}
    \label{tab:datastat}
    \begin{tabular}{ccccccl}
    \toprule
      Dataset & Total QIPs & Unique queries & Unique items & Class 2 percentage & Class 1 percentage & Class 0 percentage \\
      \midrule
human-judged\_v1 & 6.0M  & 709K & 2.3M &  64.2\% & 19.7\% & 16.1\% \\
human-judged\_v2 & 10M &  1M    & 3.7M & 63.6\%  & 19.8\%  & 16.6\%  \\
llm-judged\_v1 & 50M & 5.2M & 15M & NA & NA & NA \\
llm-judged\_v2 & 110M  & 5.2M  & 22.6M  & NA & NA & NA \\
llm-judged\_v3 & 170M  & 5.2M & 29.4M & NA & NA & NA \\
    \bottomrule
    \end{tabular}
\end{table*}

\section{Offline Experiments}
\label{sec:experiments}

\subsection{Datasets}

\paragraph{Training Datasets}
In search ranking, people usually use either search log data \citep{Wu:Puth:Meta:2024} or human editorial judgment data \citep{Wang:Sund:2024:Impr} for training text models. While collecting logged data from search engagements and impressions can help generate large-scale training data, we observe that for most of the tail queries, engagement data is often sparse and does not always reliably indicate relevance.

We leverage human-annotated labels following 5-scale relevance guidelines (as described in Table~\ref{tab:rel_label}), to create our datasets. 
Our human labeled data is generated by manually assessing the top k (usually 10) items for a set of sampled queries. We refer to these as \textit{human-judged} datasets. 
Over time, the size of the human-judged datasets increase as more data is annotated. 
We version our human-judged datasets with higher version indicating addition of more annotated data. 
We conduct experiments using two such versions shown in Table~\ref{tab:datastat}. 
When training our previous production model and teacher models, we only had access to human-judged\_v1 dataset.

To improve knowledge distillation from the teacher model, 
we also create three datasets, 
where the labels are generated from the teacher model's predictions. 
We refer to these as \textit{llm-judged} datasets. 
These datasets are created by incrementally adding more samples,
with each new version incorporating additional predictions from the teacher model.
As we expand the dataset, we update the version number. 
This also allows us to assess the impact of scaling the training data
on the student model's performance. 
All datasets are summarized in Table~\ref{tab:datastat}.

\paragraph{Test Data}

We evaluate the models using a golden test dataset 
consisting of 2,354 queries randomly selected 
from the tail segment of our search traffic. 
For each query, we sample the top 10 products and, 
an additional 10 random products from the remaining results, if available. 
This results in a total of 32,586 QIPs in the golden dataset. 
Note that the golden test data was generated after all
human evaluation data (used for training data) was collected, 
ensuring no risk of data leakage. 
While 32,000+ QIPs might seem small for a test set, 
our analysis shows that offline metrics
from this golden dataset align closely with
online human evaluations across multiple features. 
This consistency demonstrates
that the offline metrics from the golden test data are a reliable
indicator of model performance.

\subsection{Experiment Setup}

\subsubsection{Model Implementation}

We use a cross-encoder based LLM model as the teacher model 
(with the architecture described in Section~\ref{sec:LLM}), 
and fine-tune two 7B pre-trained language models:
Llama2-7B and Mistral-7B-v0.1. 
Both models are initialized from Hugging Face checkpoints and 
fine-tuned on the human-judged data \textit{human-judged\_1}.
The model training is conducted on 8 NVIDIA 40GB A100 GPUs using the AdamW optimizer. 
Building on the work of \citet{Mehr:Moha:Larg:2024} on fine-tuning LLMs 
with Walmart's human-annotated data,
we adopt the hyperparameters identified as optimal in their study. 
Specifically, we set the LoRA rank to 256, as increasing it further 
did not yield performance gains for datasets of size around $10^6$ \citep{Mehr:Moha:Larg:2024}. 
We also set the LoRA $\alpha$ to 128 and the dropout rate to 0.05.
For the learning rate, we experiment with values of $1e-4$ and $1e-5$ 
using a decay scheduler, 
and find that $1e-5$ produced the best results.

We use the $BERT_{\text{Base}}$ model with a cross-encoder architecture as the student model
(as described in Section~\ref{sec:student}). 
The student model is trained using the same learning rate of $1e-5$.

\subsubsection{Online Serving}
We deploy our Bert-based student model, $XEv2$, into production.
The output from the $XEv2$ model for a given (query, item) pair - 
referred to as the $XEv2$ feature score - 
is used as a feature in our ranking model. 
To reduce latency, we implement two optimizations: 
1. We precompute $XEv2$ feature scores for frequent query-item pairs 
as query signals through an offline pipeline.
2. We cache the $XEv2$ feature scores to further minimize real-time computation. 
For query-item pairs not present in the cache or query signal, 
the $XEv2$ model is invoked at run-time. 
To further enhance the runtime performance of the cross encoder model, 
we employ techniques like operator fusion and intermediate representations 
using the TensorRT framework~\citep{vanholder2016efficient}.

\subsubsection{Evaluation Metrics}
The models are evaluated on two tasks to measure semantic relevance: 
a classification task (determining whether a product is an exact match) and a ranking task. 
We use the following two offline metrics to assess performance:

\begin{itemize}
    \item \textbf{Recall at a fixed precision}: This metric evaluates performance on the classification task. 
    It measures the percentage of relevant products retrieved at specific precision levels
    for our  golden test dataset. 
    In our work, we fix the precision at 90\% and 95\%. 
    We denote recall at these precision levels as R@P=95\% and R@P=90\%, respectively.
    \item \textbf{Normalized Discounted Cumulative Gain (NDCG)}: 
    This metric evaluates performance on the ranking task. 
    Here NDCG@5 and NDCG@10 are used to measure
    ranking quality for the top 5 and top 10 documents, respectively.
\end{itemize}

\mycomment{
\begin{table*}[ht]
    \caption{will drop this table: Offline Experiments results for LLM teacher models}
    \label{tab:experiments_teacher}
    \begin{tabular}{ccccccl}
    \toprule
      Model  & Train Data & Starting checkpoint & R@P=95\% & R@P=90\% & NDCG@5 & NDCG@10 \\
    \midrule
    $XEv1$ & human-judged\_1 & Walmart pre-trained Bert-Base  & 0.701  & 0.844 & 0.7877 & 0.7406 \\
    Llama2-7B & human-judged\_1 & Published model & 0.784 & 0.891 & 0.7957 & 0.7465 \\
    Mistral 7B & human-judged\_1 & Published model & 0.794 & 0.899 & 0.7982 & 0.7482 \\
    \bottomrule
    \end{tabular}
\end{table*}
}
\begin{table*}[ht]
    \caption{Offline Experiment results for LLM teacher models}
    \label{tab:experiments_teacher_relative}
    \begin{tabular}{ccccccl}
    \toprule
      Model  & Train Data & Starting Checkpoint & R@P=95\% & R@P=90\% & NDCG@5 & NDCG@10 \\
    \midrule
    $XEv1$ (Baseline) & human-judged\_1 & Walmart pre-trained Bert-Base  & 0\% & 0\% & 0\% & 0\% \\
    Llama2-7B & human-judged\_1 & Hugging Face checkpoint & +11.84\%  & +5.57\% & +1.02\% & +0.8\%  \\
    Mistral-7B & human-judged\_1 & Hugging Face checkpoint &  +13.27\% & +6.52\% & +1.33\% & +1.03\% \\
    \bottomrule
    \end{tabular}
\end{table*}

\mycomment{
\begin{table*}[ht]
    \caption{will drop this table: Experiments results for Bert-Base models}
    \label{tab:experiments_student}
    \begin{tabular}{cccccccl}
    \toprule
      Model & Train Data & Loss & Starting checkpoint  & R@P=95\% & R@P=90\% & NDCG@5 & NDCG@10 \\
    \midrule
   $XEv1$ & human-judged\_1 & multi-class CE & Walmart pre-trained  & 0.701  & 0.844 & 0.7877 & 0.7406 \\
   $XEv1.1$ & human-judged\_2 + Hard Negs & soft-target CE & Walmart pre-trained & 0.771 & 0.881 &    0.7938 & 0.7455 \\
   $XEv1.2$ & llm-judged\_1 & KL loss & $XEv1.1$ & 0.685 & 0.861 & 0.7969 & 0.7472 \\
   $XEv1.3$ & llm-judged\_1 & Margin MSE & $XEv1.1$ & 0.789 & 0.894 & 0.7975 & 0.7474 \\
   $XEv1.4$ & llm-judged\_2 & Margin MSE & $XEv1.1$ & \textbf{0.794} & 0.894 & 0.7984 & \textbf{0.7483} \\
   $XEv2$ & llm-judged\_3 & Margin MSE & $XEv1.1$ & 0.791 & \textbf{0.895} & \textbf{0.7995} & \textbf{0.7483} \\
    \bottomrule
    \end{tabular}
\end{table*}
}

\begin{table*}[ht]
    \caption{Offline Comparison of Bert-Base models}
    \label{tab:experiments_student_relative}
    \begin{tabular}{cccccccl}
    \toprule
      Model & Train Data & Loss Function & Starting Checkpoint  & R@P=95\% & R@P=90\% & NDCG@5 & NDCG@10 \\
    \midrule
   $XEv1$ (baseline) & human-judged\_1 & multi-class CE & Walmart pre-trained  & 0\% & 0\% & 0\% & 0\% \\
   $XEv1.1$ & human-judged\_2 & soft-target CE & Walmart pre-trained & +9.99\% & +4.38\% & +0.77\% & +0.66\% \\
   $XEv1.2$ & llm-judged\_1 & CE & $XEv1.1$ & -2.28\% & +2.01\% & +1.16\% & +0.89\% \\
   $XEv1.3$ & llm-judged\_1 & Margin MSE & $XEv1.1$ & +12.55\% & +5.92\% & +1.24\% & +0.92\% \\
   $XEv1.4$ & llm-judged\_2 & Margin MSE & $XEv1.1$ & \textbf{+13.27\%} & +5.92\% & +1.36\% & \textbf{+1.04\%} \\
   $XEv2$ & llm-judged\_3 & Margin MSE & $XEv1.1$ & +12.84\% & \textbf{+6.04\%} & \textbf{+1.5\%} & \textbf{+1.04\%} \\
    \bottomrule
    \end{tabular}
\end{table*}

\subsection{Experiments Results}

For both teacher and student model experiments,
we use the cross-encoder based $BERT_{\text{Base}}$ ($XEv1$, 
as shown in Tables~\ref{tab:experiments_teacher_relative} and~\ref{tab:experiments_student_relative}) 
from the previous production systems as our baseline.
This choice is driven by the fact 
that live experiments (see Section~\ref{sec:online_experiment}) consistently compare the performance of the new models against the existing production model. 

\paragraph{Experiments with Teacher Models}
Table~\ref{tab:experiments_teacher_relative} compares the performance of Llama2-7B and Mistral-7B 
to the baseline $BERT_{\text{Base}}$ model on the golden test data.
All three models are fine-tuned with the human-annotated dataset \textit{human-judged\_1}.
Specifically, the $BERT_{\text{Base}}$ $XEv1$ model is fine-tuned starting from a Walmart pre-trained $BERT_{\text{Base}}$ checkpoint, which was trained on a masked language modeling task and a prediction task \citep{Mehr:Moha:Larg:2024}.
In contrast, both the Llama2-7B and Mistral-7B models are fine-tuned starting from Hugging Face checkpoints. 
Compared to the baseline $XEv1$, both Llama2-7B and Mistral-7B achieve competitive performance 
on both classification and ranking tasks. 
Notably, Mistral-7B outperforms the baseline with +13.27\% in R@P=95\% and +6.52\% in R@P=90\%. 
Its NDCG@5 and NDCG@10 scores also improve by +1.33\% and +1.03\% respectively, 
showing better performance on the ranking task as well. 
Based on the results in Table~\ref{tab:experiments_teacher_relative}, 
since Mistral-7B performs the best as a teacher model, 
we use its predictions to label a large-scale augmented dataset for knowledge distillation.

For the student models, we train multiple $BERT_{\text{Base}}$ models 
on the generated large-scale datasets 
and conduct an ablation study to examine the impact of each factor
on model performance. 
This process allows us to address the following research questions
and share insights with machine learning practitioners.



\subsubsection{RQ1: What is the most effective knowledge distillation loss 
when both the student and teacher models use the same architecture?}
Previous work by \citet{hofstatter2020improving}
shows that different model architectures, 
such as cross-encoder based BERT, 
bi-encoder based BERT, and ColBERT, 
produce output scores with varying magnitudes.
Their proposed Margin-MSE loss effectively adapts to 
these differences in score distributions.

In our case, both the student model and teacher models 
use the cross-encoder architecture, 
so we are interested in whether focusing on the margin between two items 
remains an effective approach for knowledge distillation.
We aim to determine the most effective approach for knowledge distillation in this context. 
We compare our extended Margin-MSE loss 
(described in Section~\ref{sec:student})
with a pointwise cross-entropy loss,
where the teacher model's prediction scores serve as the target labels.
The pointwise cross-entropy loss is defined as follows:
\begin{equation*}
   \mathcal{L}_{student'}(q) = \sum_{d \in \mathbf{D}_q} [-t_{q,d} \cdot \log s(q,d)-(1-t_{q,d}) \cdot \log(1 - s(q,d))].
\end{equation*}

We validate our strategy with an ablation study comparing two knowledge distillation losses. 
As shown in Table~\ref{tab:experiments_student_relative}, 
Model $XEv1.2$ and $XEv1.3$ were both fine-tuned on the same dataset,  \textit{llm-judged\_1},
labeled by the Mistral-7B teacher model.
The only difference between these two models is the choice of 
knowledge distillation loss. 
The results show that the Margin-MSE loss significantly outperforms the cross-entropy loss
across both precision/recall and NDCG metrics. 
Notably, we observe substantial improvements in precision/recall, 
with relative gains of +14.83\% at R@P=95\% and +3.91\% at R@P=90\%. 

\subsubsection{RQ2: Does large-scale unlabeled training data facilitate knowledge distillation from the teacher model to the student model?}
Using the extended Margin-MSE as the knowledge distillation loss,
we fine-tune $BERT_{\text{Base}}$ models ($XEv1.3$, $XEv1.4$, $XEv2$) on LLM-labeled datasets of increasing sizes.
As the dataset size grows from 50M to 110M and then to 170M, 
we observe consistent improvements in NDCG metrics, 
while performance on precision/recall metrics appears to plateau
at higher data volume (see Table~\ref{tab:experiments_student_relative}).
Overall, as the dataset expands, the student model continues to
show significant performance gains,
highlighting the effectiveness of training with large-scale teacher-generated labels.

\subsubsection{RQ3: Can a 110M-parameter student model achieve 
performance comparable to a 7B-parameter teacher LLM 
when trained on a large-scale unlabeled dataset?}

When compared to the Mistral-7B teacher model, $XEv1.4$ achieves identical performance on R@P=95\% (+13.27\% for both models) but shows a slightly lower R@P=90\% (+5.92\% for $XEv1.4$ vs. +6.52\% for the teacher model). However, $XEv1.4$ slightly outperforms the teacher model on NDCG metrics, with +1.36\% for NDCG@5 (vs. +1.33\% for the teacher) and +1.04\% for NDCG@10 (vs. +1.03\% for the teacher).
When comparing $XEv2$ to the teacher model, 
$XEv2$ shows notable improvements in NDCG metrics,
with +1.5\% for NDCG@5 (vs. +1.33\% for the teacher)
and +1.04\% for NDCG@10 (vs. +1.03\% for the teacher).
$XEv2$ slightly lags behind on precision/recall metrics though,
achieving +12.84\% for R@P=95\%  (vs. +13.27\% for the teacher)
and +6.04\% for R@P=90\% (vs. +6.52\% for the teacher).

Overall, despite being more than 60 times smaller than the teacher model,
the student model demonstrates comparable performance, 
proving that it effectively learns from the teacher.
This suggests that, in this context, 
additional model capacity may not be necessary for successful knowledge distillation.

\subsubsection{RQ4: Can the student model effectively learn from the teacher model, 
with fewer input features?}
Item descriptions have been shown to enhance performance
in multi-billion parameter LLMs \citep{Mehr:Moha:Larg:2024}.
We include the item description in the input for the teacher LLM.
However, due to the long text and latency requirements in production,
we exclude it from the student model.
Despite removing the item description from the student model's input,
the student model still achieves performance comparable to that of the teacher model.
This also demonstrates the effectiveness of knowledge distillation,
even when the student model has less input data.

\section{Online Experiments}
\label{sec:online_experiment}

We deployed the enhanced $XEv2$ feature into production for our online experiments. In these experiments, the $XEv2$ score for a query-item pair
serves as a feature in our ranking model. 
The control model uses the existing production setup with the $XEv1$ feature, 
while the variation model incorporates the new $XEv2$ feature in place of $XEv1$.

\subsection{Manual Evaluation Results}

\begin{table}[ht]
    \caption{Human evaluation of top-10 rankings for tail queries by the new model}
    \label{tab:online_eval}
    \begin{tabular}{cl}
    \toprule
       NDCG@5 Lift (P-value) & NDCG@10 Lift (P-value) \\
      \midrule
        +1.07\%  (0.00) & +0.87\% (0.01) \\
    \bottomrule
    \end{tabular}
\end{table}

We evaluate the performance of the proposed $XEv2$ by 
having human assessors evaluate the top-10 ranking results 
generated by the model, with $XEv2$ as a feature, compared to 
the existing production model that uses $XEv1$.
Queries are randomly selected from the tail segment of our search traffic. 
For each query, human assessors are presented with the product image, title, price and a link to the product on the Walmart website. 
They rate the relevance of each product on a 5-point scale, 
while NDCG metrics are calculated based on a mapped 3-class scale, in accordance with our production requirements. 
As shown in Table~\ref{tab:online_eval}, 
the new $XEv2$ feature significantly improved relevance for tail queries,
with notable lifts in both the top 5 and top 10 rankings. 
It is crucial to underscore the significance of this improvement 
given the robustness of our baseline and the maturity of our existing search ranking system. 
The search ranking system has already been optimized over time, 
making any incremental enhancement a noteworthy achievement. 
Therefore, it should be viewed as a substantial uplift.

\subsection{Interleaving Results}

\mycomment{
\begin{table}[ht]
    \caption{Interleaving results on the top-40 ranking for the proposed architecture}
    \label{tab:online_interleaving}
    \begin{tabular}{lc}
    \toprule
    & ATC-lift (P-value) \\
    \midrule
    Set-based & 0.06\% (0.00) \\
    Rank-based & 0.41\% (0.00) \\
    Rank-blocked-based & 0.21\% (0.00) \\
    \bottomrule
    \end{tabular}
\end{table}
}

Interleaving \citep{joachims2003evaluating} is an online evaluation method 
used to compare user engagement performance 
between our proposed model and the previous production model at Walmart.
During the 14-day interleaving period, 
each user is shown a mix of ranking results 
from both the control and variation models.
The metric measured is ATC@40, the count of add-to-carts (ATC) 
within the top 40 positions for both the control and variation models. 
The results show an ATC-lift of 0.06\% with a P-value of 0.00 on the top-40 ranking,
indicating a significant improvement in user engagement performance.


\subsection{Online AB Test}
\begin{table}[ht]
    \caption{AB test results on e-commerce engagement metrics 
    for the proposed architecture}
    \label{tab:online_AB}
    \begin{tabular}{lc}
    \toprule
    & Improvement \\
    \midrule
    ATC Rate per Session & +0.8\% \\
    ATC Rate per Visitor & \textbf{+1.2\%} \\
    Search Session Abandonment Rate & \textbf{-0.6\%} \\
    Clicks Needed before an ATC event & \textbf{-1.1\%} \\
    \bottomrule
    \end{tabular}
    \begin{tablenotes}
    \footnotesize 
    \item Note: Bold values indicate statistically significant results (p<0.05). 
    \end{tablenotes}
\end{table}

We conduct an online AB test using production tail traffic from August 26 to September 9, 2024 (two weeks). 
Our proposed architecture 
shows statistically significant improvements 
across multiple key engagement metrics over the previous production model. 
As reported in Table 8, the new architecture increases the average ATC rate per session by 0.8\% 
and the ATC rate per visitor by 1.2\%. 
Additionally, it reduces search session abandonment rate by 0.6\% 
and lowers the average number of clicks required before an ATC event by 1.1\%. 
These positive results lead to the successful deployment of 
the student model in production.
It now handles the traffic for tail queries, 
significantly improving the customer experience. 

\section{Discussion and Future Work}
\label{sec:conclusion}

In this work, we present a knowledge distillation framework
tailored for search re-ranking tasks.
Our results show that the student model can be significantly improved
through knowledge distillation using the Margin-MSE loss,
as compared to pointwise cross-entropy loss.
This suggests that learning the margin between two items 
is more effective than optimizing pointwise loss,
even when both the student model and teacher models 
use a cross-encoder architecture. 
We also conduct an ablation study to examine the impact 
of augmented training data size on model performance.
Our results show that increasing the training data size
significantly boosts the student model's performance.
However, once the data reaches a certain size, 
performance improvements plateau.

The student model, $XEv2$, trained using this framework,
has been successfully deployed in the re-ranking stage 
of Walmart's e-commerce platform, 
resulting in notable improvements in both relevance and engagement.
This gains are validated through both offline and online evaluations. 

Despite these promising results, 
there are a few limitations in our current approach.
Due to computational and time constraints,
we did not experiment with different versions 
of human-judged data for training the teacher model.
Future work could include more experiments with the teacher model, 
which may further enhance the performance of 
both the teacher and student models.
Additionally, we did not incorporate query attributes
(e.g., query product type) as input for the teacher model,
which could provide further context and improve model performance.
Furthermore, while the current $XEv2$ model focuses solely 
on learning relevance between query-item pairs,
we plan to explore incorporating a multi-objective loss function 
that combines both relevance and engagement metrics. 
Finally, while this current $XEv2$ model is deployed for tail queries, 
we aim to expand its use to head/torso query segments in the near future.

\begin{acks}
We would like to sincerely thank Juexin Lin, Nick Rossi, and Tony Lee for providing the ANN data and for their valuable discussions. Our gratitude also goes to Ciya Liao and Matthew Mu for their useful guidance. We appreciate Mossaab Bagdouri for his help with the manual evaluations and Atul Singh for his work on implementing the query signal pipeline. Last but not least, we would like to extend our heartfelt thanks to Yuan-Tai Fu, Bhavin Madhani, and the entire engineering team. Without their support, the successful operation of the pipeline would not have been possible.
\end{acks}

\bibliographystyle{ACM-Reference-Format}
\balance
\bibliography{XE}

\end{document}